\documentclass[twocolumn,pre]{revtex4}
\usepackage[english]{babel}
\usepackage[utf8x]{inputenc}
\usepackage[pdftex]{graphicx}
\usepackage[a4paper, top=1in, bottom=1.25in, left=0.8in, right=0.8in]{geometry}
\usepackage{enumitem}
\usepackage{amsmath}
\usepackage{bm}
\usepackage{braket}
\usepackage{float}

\graphicspath{{./Figures/}}
\newcommand{\dd}[1]{\mathrm{d}#1}
\setlist{leftmargin=*}

\begin{document}

\title{Orientational ordering of point dipoles on a sphere}
\author{Andraž Gnidovec, Simon Čopar}
\affiliation{Faculty of Mathematics and Physics, University of Ljubljana, SI-1000, Ljubljana, Slovenia}

\begin{abstract}
Arrangement of interacting particles on a sphere is historically a well known problem, however, ordering of particles with anisotropic interaction, such as the dipole-dipole interaction, has remained unexplored. We solve the orientational ordering of point dipoles on a sphere with fixed positional order with numerical minimization of interaction energy and analyze stable configurations depending on their symmetry and degree of ordering. We find that a macrovortex is a generic ground state, with various discrete rotational symmetries for different system sizes, while higher energy metastable states are similar, but less ordered. We observe orientational phase transitions and hysteresis in response to changing external field both for the fixed sphere orientation with respect the field, as well as for a freely-rotating sphere. For the case of a freely rotating sphere, we also observe changes of the symmetry axis with increasing field strength.
\end{abstract}

\maketitle

\section{Introduction}

Minimal energy distributions of interacting particles on a sphere is a well known problem both for its historical significance and contemporary relevance. Since J. J. Thomson proposed his model of an atom in 1897 and in turn sought configurations with minimal energy for $N$ same-charged particles on the surface of a sphere \cite{Thomson}, the problem has been generalized to different interparticle interactions, most notably, different long-range power-law \cite{Bowick1} and logarithmic \cite{Saff1997} interactions, Tammes problem of the packing of hard circles \cite{Tammes}, and arrangement of connected charges \cite{Slosar}. Active research on different interaction potentials and geometric aspects of solutions continues to this day \cite{BowickPRL,Irvine,Wales,Wales2009,Miller}. Investigation of sphere-bound particles under effect of generalized interactions gives insights into the symmetry and geometry of the resulting organizational order, and yields descriptive models of various types of self-organized matter, such as arrangement of proteins in capsids \cite{Marzec,ZandiR_ProcNatlAcadSci101_2004}, fullerene patterns in carbon clusters \cite{Kroto}, and distribution of solid particles in Pickering emulsions \cite{Chevalier}.

In contrast with the body of research on isotropic particles, ordering of discrete particles with orientation-dependent interactions on a sphere has remained relatively unexplored. Anisotropy can be a consequence of non-circular hard particles \cite{Li2013}, directed motion in dynamical systems \cite{Luca,Sknepnek,Praetorius}, short-range nearest neighbor couplings, such as approximate models of spin lattices, or in general, induced by anisotropic long-range interactions. A natural anisotropic extension of the Thomson problem, which has not been considered before, is to extend the multipolar expansion to the dipolar term, so that in addition to position, orientation of a polarization vector can be varied for each particle. When the repulsive isotropic interactions between the particles are strong enough, a restricted problem can be considered, fixing the particle positions and solving for polarization orientations that minimize the electrostatic energy of the system.

In this paper, we investigate ground state orientations of point dipoles on a sphere, positionally fixed to the Thomson lattice. Different 2D lattices of interacting dipoles have been studied already during the previous century \cite{Prakash,Zimmerman,Belobrov} and it was found that long-range nature of dipole-dipole interactions has determining effect on orientational ordering and structural phase transitions. While the topology of a sphere prevents us from fitting any regular lattice to its surface \cite{Bowick2}, these solutions can shed light on what to expect from dipole systems on a sphere and help with interpretations of the results. The paper is organized as follows. We start by defining the Hamiltonian, describe simulation methods used to find the ground states of dipole systems on a sphere, and introduce order parameters for quantitative analysis of order. We present the results of numerical simulations and show that the ground states have a macrovortex structure for all numbers of dipoles. We further study the effects of external field on dipole configurations and the resulting orientational phase transitions, both for fixed field direction and for the case of a freely rotating sphere.

\section{Simulation methods}

We consider a system of $N$ dipoles $\{\bm{p}_i\}$ fixed at points in space $\{\bm{r}_i\}$ and in the presence of an external field $\bm{H}$. Like in the standard Thomson problem, interactions between all particle pairs are taken into account. The Hamiltonian of our system is
\begin{equation}\label{hamilton}
\mathcal{H} = \sum_{\substack{i, j = 1 \\ i\neq j}}^{N} \frac{\bm{p}_i \cdot \bm{p}_j - 3(\bm{p}_i \cdot \hat{\bm{r}}_{ij})(\bm{p}_j \cdot \hat{\bm{r}}_{ij})}{r_{ij}^3} -\bm{H}\cdot \sum_{i=1}^N \bm{p}_i,
\end{equation}
where the first term represents dipole-dipole contribution to energy in units of $1/4\pi\epsilon_0$ and the second term describes interaction of dipoles with the external field. We denoted $\bm{r}_{ij} = \bm{r}_j - \bm{r}_i$ and hat over vector represents unit vector in the same direction. The positional order is determined by the solution of the Thomson problem as one of the representative uniform point distributions on a sphere.  Orientations of dipoles are parametrized by two angles, azimuthal angle $\phi_i$ and polar angle $\theta_i$, in their respective local coordinate frames. We solve the problem of energy minimization numerically. Considering the size of the systems we want to explore ($N<100$---higher values show no new structures compared to lower $N$), we choose BFGS minimization algorithm as both an effective and time efficient method to find ground states and other stable configurations in zero external field.
The results of the minimization can depend on the initial configuration -- the system can relax to a metastable configuration (local minimum of the Hamiltonian) that the minimization algorithm cannot escape, which results in a large number of end states with different degrees of ordering. We perform the minimization repeatedly from different random initial conditions to ensure finding the global minimum.

\begin{figure}[!ht]
	\centering
	\includegraphics[width=\linewidth]{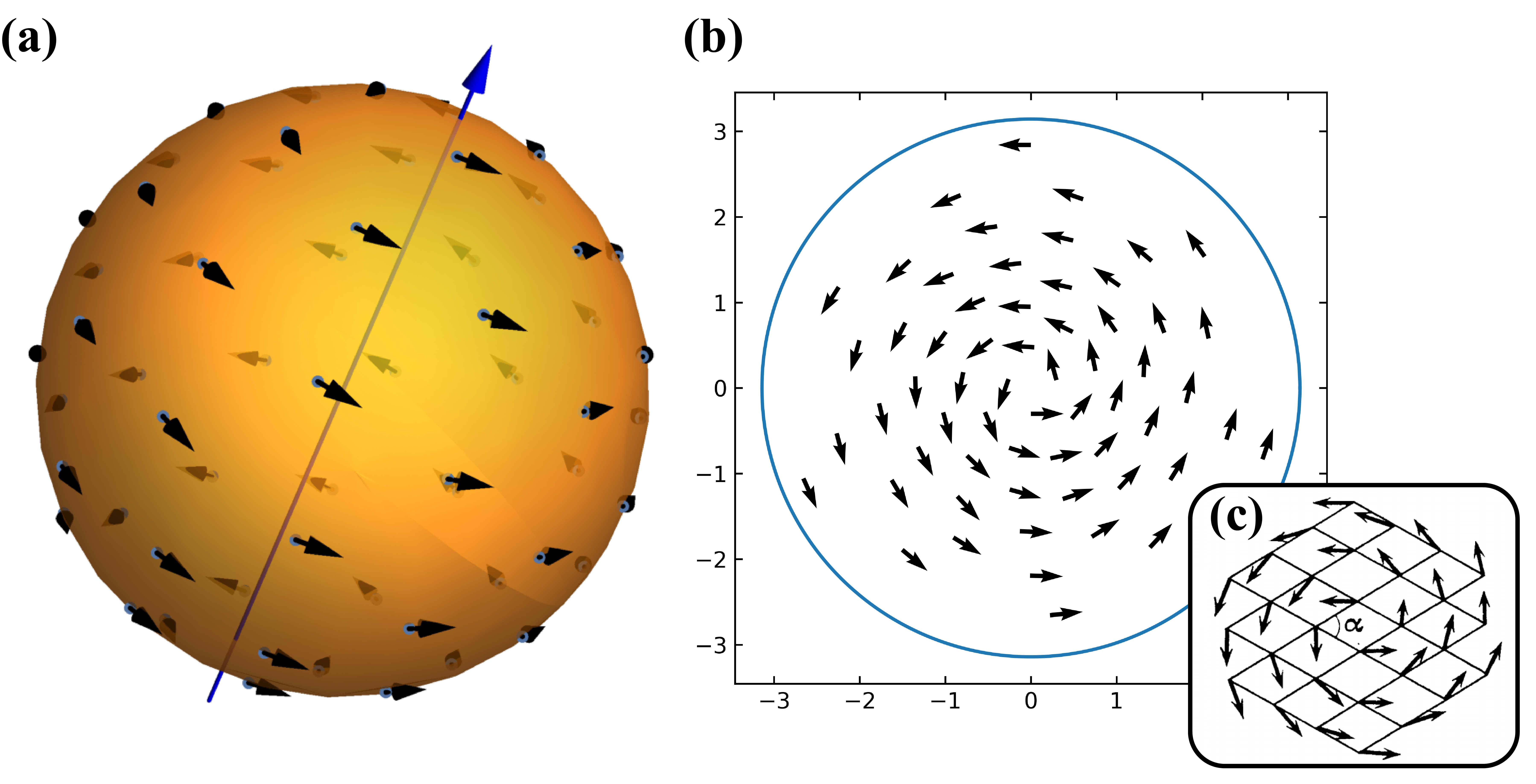}
	\caption{\textbf{(a)} 3D visualization of the ground state configuration for $N=60$. Macrovortex orientation is formed where the orientation of dipole moments is characterized by the direction of angular momentum $\bm{\Gamma}$, here represented by blue arrow. \textbf{(b)} The same configuration showed in azimuthal projection. The direction of $\bm{\Gamma}$ defines the north pole of the sphere and lies in the center of the diagram. The entire blue circle represents the south pole. \textbf{(c)} Ground state of dipole system on hexagonal lattice is also a macrovortex (adapted from \cite{Belobrov}).}
	\label{fig:vrtinci}
\end{figure}

To quantitatively measure the order so that we can compare various configurations, we need to find appropriate order parameters. The choice of parameters depends on expected ordering of dipoles. Consider results of the energy minimization for 2D lattice dipole systems. For square and hexagonal lattice, the ground state is infinitely degenerate and periodic \cite{Prakash}.
For angles of rhombicity between  $\alpha=50\text{°}$ and $\alpha=75\text{°}$, which include the special case of the triangular lattice at  $\alpha=60\text{°}$, the ground state is found to be a macrovortex \cite{Belobrov}. The Thomson lattice locally resembles the triangular lattice -- except for the lattice defects, topologically required by the Euler characteristic of the sphere, most vertices have six nearest neighbors. 
Our calculations indeed show that the macrovortex structure is also the ground state of the dipole system on a sphere with positional order fixed to Thomson positions. Figure~\ref{fig:vrtinci}a shows the ground state configuration for $N=60$. The macrovortex structure stands out even more in the azimuthal projection in Fig.~\ref{fig:vrtinci}b, shown in comparison to the macrovortex on 2D triangular lattice (Fig.~\ref{fig:vrtinci}c). The behavior of 2D lattice dipole systems in external field also gives indication of expected field response for dipole systems on a sphere. For example, a hexagonal lattice shows a discontinuous orientational phase transition in the external field \cite{Zimmerman}, leading us to expect similar behavior for spherical dipolar systems.

To determine axis of rotation and quantify the macrovortex nature of the ground states, we define angular momentum
\begin{equation}
\bm{\Gamma} = \frac{1}{N} \sum_i \hat{\bm{r}}_i \times \hat{\bm{p}}_i.
\end{equation}
The amplitude $\Gamma$ gives us information on the intensity of this circulation.
To explore the response of the system to external magnetic field $\bm{H}$, we further define magnetization
\begin{equation}\label{magnetizacija}
\bm{M}=\frac{1}{N}\sum_{i=1}^N \bm{p}_i
\end{equation}
and susceptibility
\begin{equation}\label{susceptibilnost}
\chi = \frac{\dd M_{\parallel}}{\dd H},
\end{equation}
where $M_{\parallel}$ denotes the component of magnetization in the direction of $\bm{H}$. For calculations of response to external field, where the system must stay in the same local minimum during changes of the field and starting from the chosen stable configuration, we use a simple relaxation method (gradient descent) instead of a global minimization algorithm. Our Hamiltonian is a quadratic form and relaxation therefore reduces to iterative application of a linear transformation and renormalization of dipole moments. As we gradually increase the amplitude of external field, we calculate magnetization and susceptibility at each step. This continuity of states is also important in measuring the hysteresis response. 

\section{Results}

In this section we present our simulation results, first focusing on analyzing different stable states found. We later explore the effects of external magnetic field on these states both for fixed direction of the field as well as for the case of a freely-rotating sphere. 

\subsection{Stable configurations}

\begin{figure}[!ht]
	\centering
	\includegraphics[width=0.9\columnwidth]{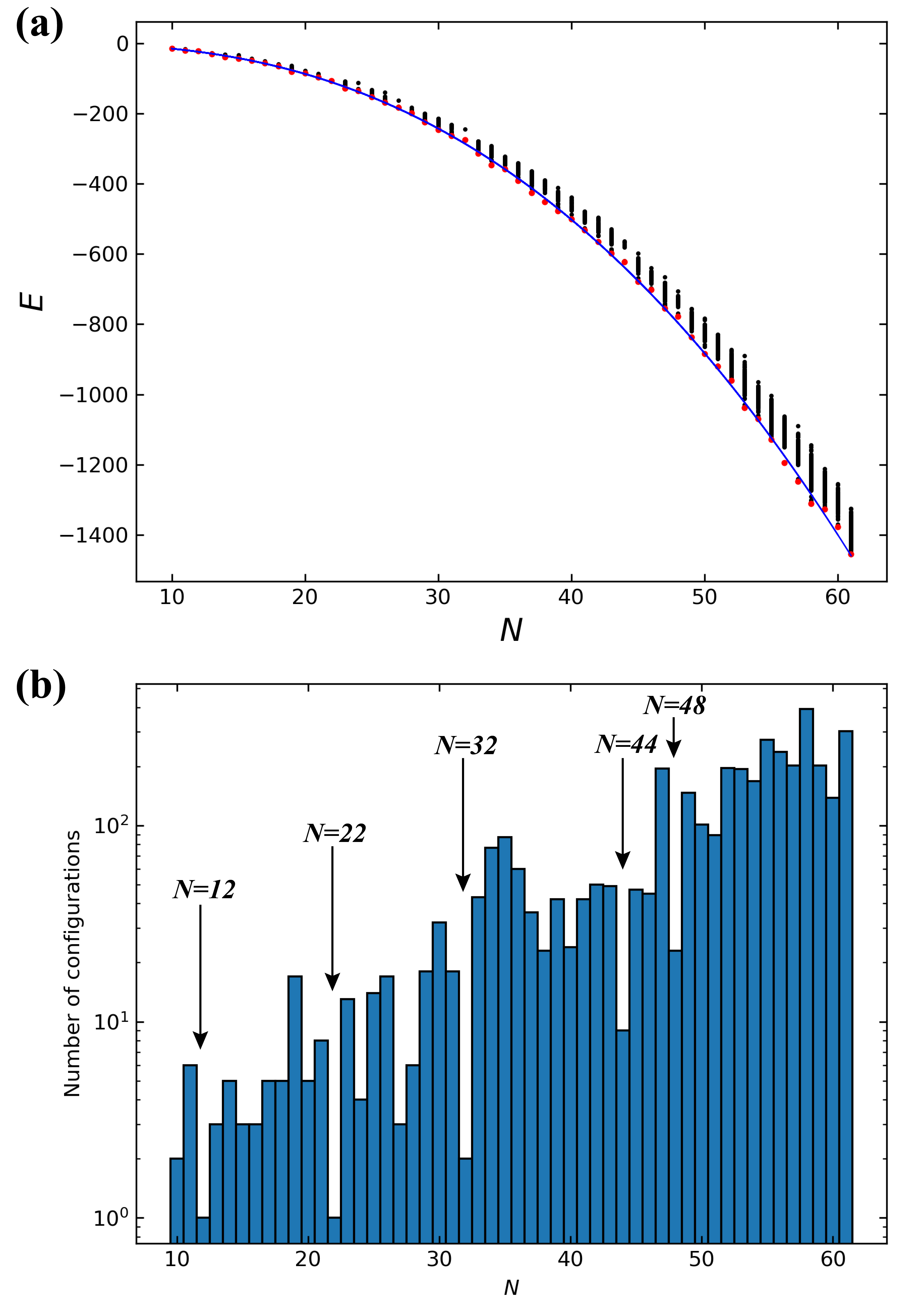}
	\caption{Results of 1000 minimization simulations for all $N$ between 10 and 61. \textbf{(a)} Energies of all found configurations. Ground state energies are represented by red dots while excited state energies are shown in black. We also fit power law curve to ground state energies and determine scaling as $E_0 \sim N^{2.53}$. \textbf{(b)} Number of different configurations found for each $N$ shows exponential growth trend (note the logarithmic scale on $y$-axis). Arrows indicate $N$ that correspond to symmetric lattices, which show much smaller number of local minima.}
	\label{fig:konfiguracije}
\end{figure}

We perform 1000 minimization simulations at each $N$ to examine the dependence of ground state energy and number of all states found on the number of dipoles (Fig~\ref{fig:konfiguracije}). Initial simulations were performed with no restrictions to dipole orientations. Nevertheless, the results show that in all obtained configurations dipoles orient tangent to the surface of the sphere. This is in line with expectations as it can be shown that in the ground state configuration of long-range interacting dipole system, dipoles orient in the way that minimizes bulk magnetization~$\bm{M}$ \cite{Belobrov}. On a sphere this means ordering tangent to the surface. Subsequent minimization simulations were therefore performed with dipoles constrained to lying in their respective tangent planes which reduces the dimension of the problem from $2N$ to $N$ and in turn decreases calculation times. 

We find that the ground state energy decreases monotonously and can be fitted by a power law curve $E_0\propto N^{2.53}$. This exponent value is close to the estimate $5/2$ we get by taking into account scaling of the distance between dipoles as $r\propto N^{-1/2}$ and energy as $E\propto N/r^3$. The dependence of the number of different configurations (local minima) on $N$ is more complicated and strongly connected to the symmetry of positional order.
We notice the trend of exponential increase of found local minima with $N$ (see Fig.~\ref{fig:konfiguracije}b), which is expected as higher number of dipoles on the sphere allows for more stable but frustrated local configurations that prevent global ordering. Systems with low number of found states can be linked to high symmetry of positional order. The cases that stand out the most are $N=12$ and $N=22$ with only one configuration, $N=32$ with 2 configurations and $N=44$ with 9 configurations, with icosahedral, tetrahedral, icosahedral and octahedral positional order symmetries, respectively. Similarly, we also find lower number of states for other high symmetry configurations compared to surrounding values of $N$.

\begin{figure}[!ht]
	\centering
	\includegraphics[width=\linewidth]{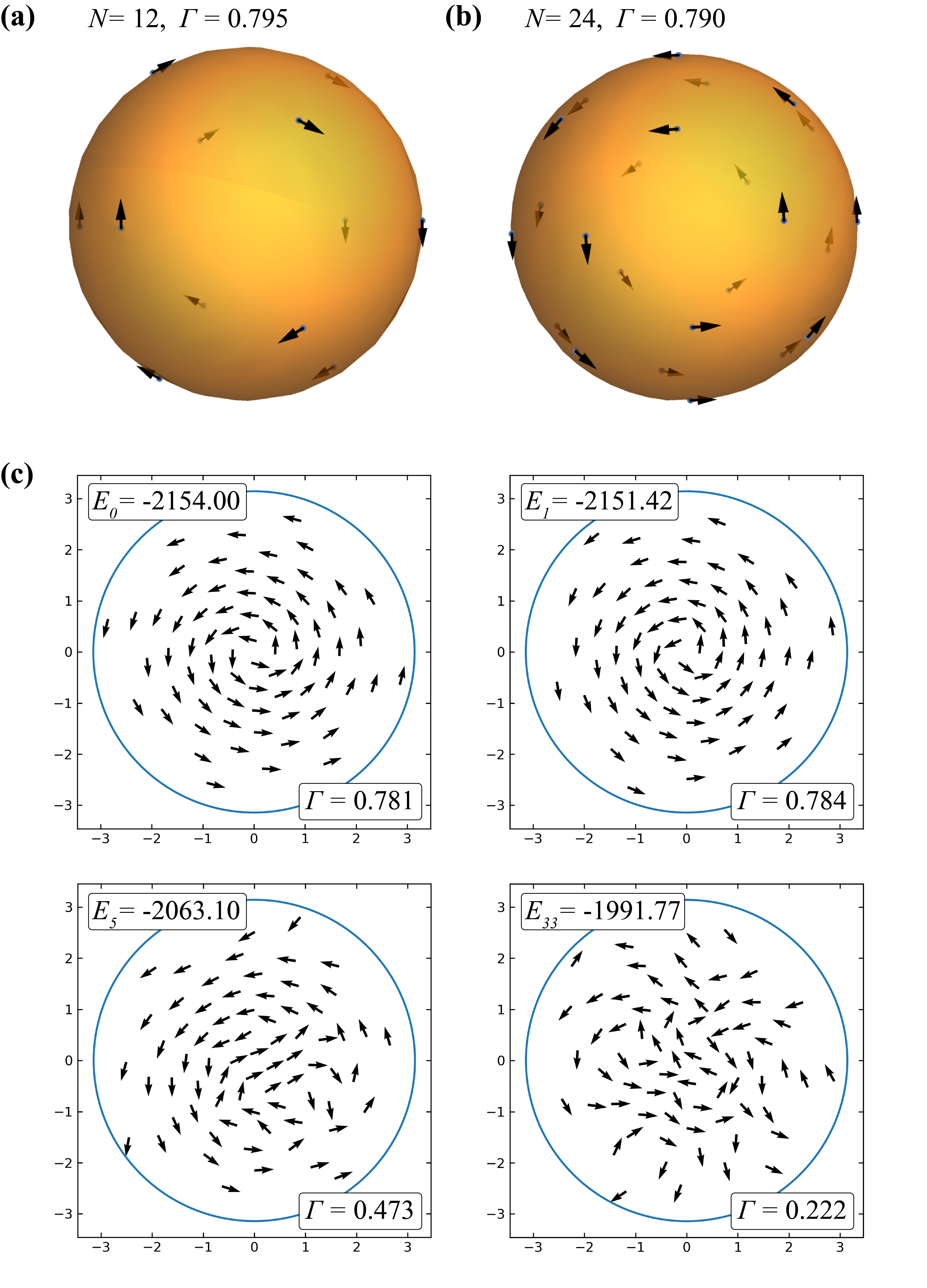}
	\caption{Visualization of some configurations at different $N$. Angular momentum amplitudes characterizes the degree of order in the system with values close to $\pi/4\approx0.785$ indicating macrovortex structure and lower values related to less ordered configurations. \textbf{(a)} Ground state configuration for $N=12$ shows $C_3$ symmetry. All dipoles are also aligned in planes perpendicular to the direction of $\bm{\Gamma}$. \textbf{(b)} Ground state configuration for $N=24$ belongs to $C_4$ point group. \textbf{(c)} Azimuthal projections for ground state and three excited states of $N=72$ case. The first two states show different macrovortex configurations---the ground state configuration has a two-fold symmetry axis while the first excited state has a three-fold axis. We also show one of partially ordered states where only approximately half of all dipoles are oriented in a vortex while the other are disordered, and one of disordered states.}
	\label{fig:vizualizacija}
\end{figure}

We now examine individual cases of stable structures. First we look at the ground state for $N=12$ (Fig.~\ref{fig:vizualizacija}a). Macrovortex configuration is formed which reduces the symmetry of the solution from icosahedral for positional order to the point group $C_3$. The direction of angular momentum $\bm{\Gamma}$ corresponds to the three-fold rotation axis of the configuration. All of the dipoles lie on four planes perpendicular to $\bm{\Gamma}$ with their dipole moments also parallel to this planes. The amplitude of angular momentum is $\Gamma=0.795$ which is close to the analytical estimate of $\pi/4\approx0.785$ for continuous distribution of dipoles arranged in a macrovortex (the deviation here is the consequence of discretization). Also possible are configurations with 4-fold symmetry axis, for instance the ground state for $N=24$ shown in Fig.~\ref{fig:vizualizacija}b. The amplitude of $\bm{\Gamma}$ is again close to the analytical estimate, a characteristic that emerges also for macrovortex states at other $N$ and can therefore be used as indicator for the degree of ordering, with less ordered configurations described by lower value of $\Gamma$.

Finally, we look at four different configurations of $N=72$ dipoles where positional order again has icosahedral symmetry (Fig.~\ref{fig:vizualizacija}c). As expected, the number of all states found in 1000 minimization simulations is 35, low for a system of this size. We use azimuthal projection for better visualization. Ground state (top left panel) and first excited state (top right panel) show the formation of two distinct macrovortex structures with the former being less symmetric (two-fold symmetry) than the other (three-fold symmetry). As an illustration of possible partially ordered and disordered states that emerge for higher numbers of dipoles, we also show configurations for fifth (bottom left panel) and thirty-third (bottom right panel) excited state. The amplitude of angular momentum decreases as the macrovortex ordering disappears. In the first case, approximately one half of dipoles is already forming a macrovortex, however, the other dipoles are locked in a different configuration (local energy minimum). Similarly, the last case shows the formation of dipole strings that are also metastable.

In general, we find macrovortex ground states for all values of $N$. This can also be confirmed by the values of $\Gamma$ that are close to the analytical estimate of $\pi/4$ while lower values indicate less ordered configurations. One should note that in spite of macrovortex structure, it is possible that the dipole system has no exact symmetries. The showcased examples benefit from high positional order symmetry but even that does not guarantee symmetrical configurations of dipole states, e.g. the ground state for $N=32$ has no symmetries. On the other hand, we saw that some configurations also exhibit higher symmetries than the most common $C_3$ symmetry, for instance $C_4$ for $N=24$.

\subsection{Fixed direction of external field}

The simulations for determining the response of stable dipole configurations on a sphere in external field were performed using relaxation approach that also models the correct system dynamics under slow changes of external field. The restriction of tangent ordering is of course lifted and dipoles are again freely-rotating. First, we limit ourselves to the case with fixed direction of the field. The choice of this direction is has an important impact on the results. We choose field parallel to angular momentum $\bm{\Gamma}$ of the system which represents the characteristic direction of dipole order for each configuration. While this makes the comparison of behavior between different configurations difficult, it enables us to roughly grasp the properties of the system in external field.  

\begin{figure*}[!ht]
	\centering
	\includegraphics[width=\linewidth]{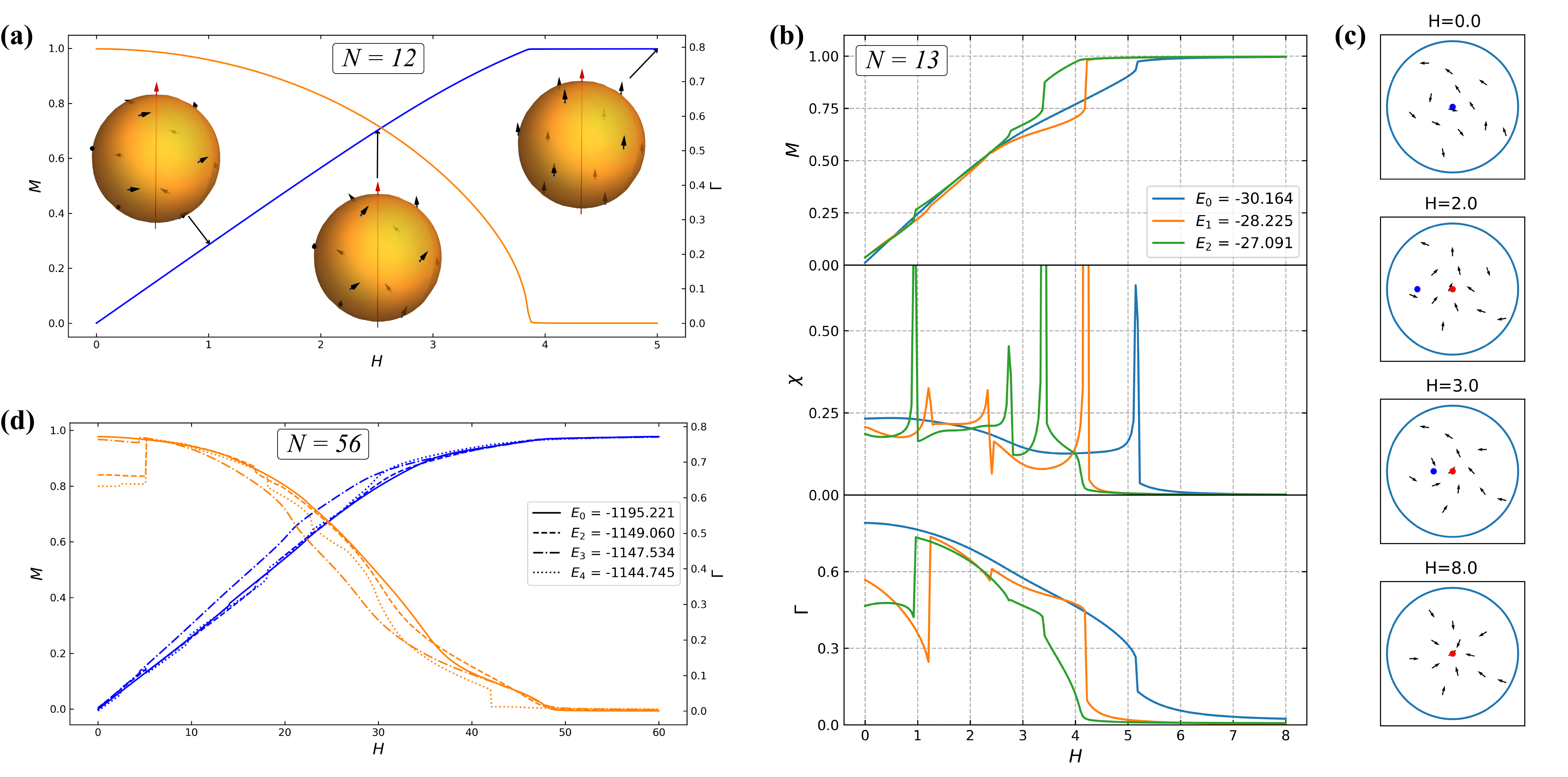}
	\caption{\textbf{(a)} Magnetization curve (blue) and amplitude of angular momentum (orange) in dependence to the amplitude of external field. As shown in complementary 3D visualizations at different values of external field amplitude ($H=1.0$, $H=2.5$ and $H=5.0$), the dipoles gradually reorient from the initial macrovortex configurations and at high amplitudes fully align with the field. The red arrow shows the direction of external field. Orientational phase transition in this case is continuous. \textbf{(b)} Magnetization, susceptibility and amplitude of angular momentum for all three stable configurations of $N=13$ dipoles. In contrast to $N=12$ case, we notice multiple discontinuous phase transitions that result in divergences in susceptibility. The jump in $\Gamma$ at first phase transition for two excited states shows that in external field, some configurations can become more ordered. \textbf{(c)} Azimuthal projection, modified to show normalized projections of dipole orientations to their respective tangent planes, of configurations for the first excited state of $N=13$ case at different filed amplitudes. We use the blue dot to show the direction of angular momentum $\bm{\Gamma}$ for depicted configuration and the red dot to show the direction of external field (aligned with initial $\bm{\Gamma}$ of the configuration). In the first frame, blue and red dot coincide. In the last frame the direction of $\bm{\Gamma}$ is not shown anymore as the amplitude is too small for the quantity to be relevant. \textbf{(d)} Magnetization and $\bm{\Gamma}$ amplitude for different configurations of $N=56$. Discontinuous phase transitions are especially noticeable for the excited states.}
	\label{fig:zunanje_polje}
\end{figure*}

We start by examining the response for the simplest case of $N=12$ (Fig.~\ref{fig:zunanje_polje}a). Magnetization increases continuously until saturation and similarly, angular momentum amplitude drops to zero. This signals that the system undergoes a second-order orientational phase transition from macrovortex to total alignment with the field. The change in dipole directions can be seen in simulation frames taken at different external field amplitudes.
Continuous orientational phase transition is not a general result and emerges for $N=12$ because of the positional order and configuration symmetry. We observe similar behavior for other configurations with $C_n$ symmetry while ground states and excited states that are not symmetric exhibit different characteristics. Figure~\ref{fig:zunanje_polje}b shows dependence of magnetization, susceptibility and amplitude of angular momentum on the amplitude of the external field for three metastable configurations for $N=13$. The ground state is a macrovortex while the excited states show non-regular ordering. We notice that all magnetization curves have at least one discontinuous jump that reflects in the divergence in susceptibility, signalling orientational phase transitions. To better understand the nature of these transitions, we look at the graphs of angular momentum amplitude where general decrease of $\Gamma$ shows that the dipoles are aligning with the field. More interesting is the discontinuous jump in $\Gamma$ at first phase transition where the system relaxes to a more ordered quasi-macrovortex structure. To better imagine how the order of the system changes with the increasing magnetic field, we use azimuthal projection to show dipole orientations at different field amplitudes. The results for the first excited state of $N=13$ are presented in Fig.~\ref{fig:zunanje_polje}c. We notice that after the first orientational phase transition (frame 2), angular momentum shifts away from the direction of the field to form a new macrovortex configuration. After the second phase transition (frame 3) the dipoles order to form a dipole string and after the third transition, dipoles align with the external magnetic field. In the last frame the direction of $\bm{\Gamma}$ is not shown anymore as the amplitude is too small for the quantity to be relevant. 

As an example of higher $N$ we show the magnetization and angular momentum amplitude curves for the ground states and three lower excited states (second, third and fourth) of $N=56$. The results are similar to ones for $N=13$ with second and fourth excited states undergoing more orientational phase transitions than the ground state. The configuration of the third excited state is also a macrovortex and the number of discontinuous transitions is therefore lower. 

\begin{figure}[!ht] 
	\centering
	\includegraphics[width=0.8\linewidth]{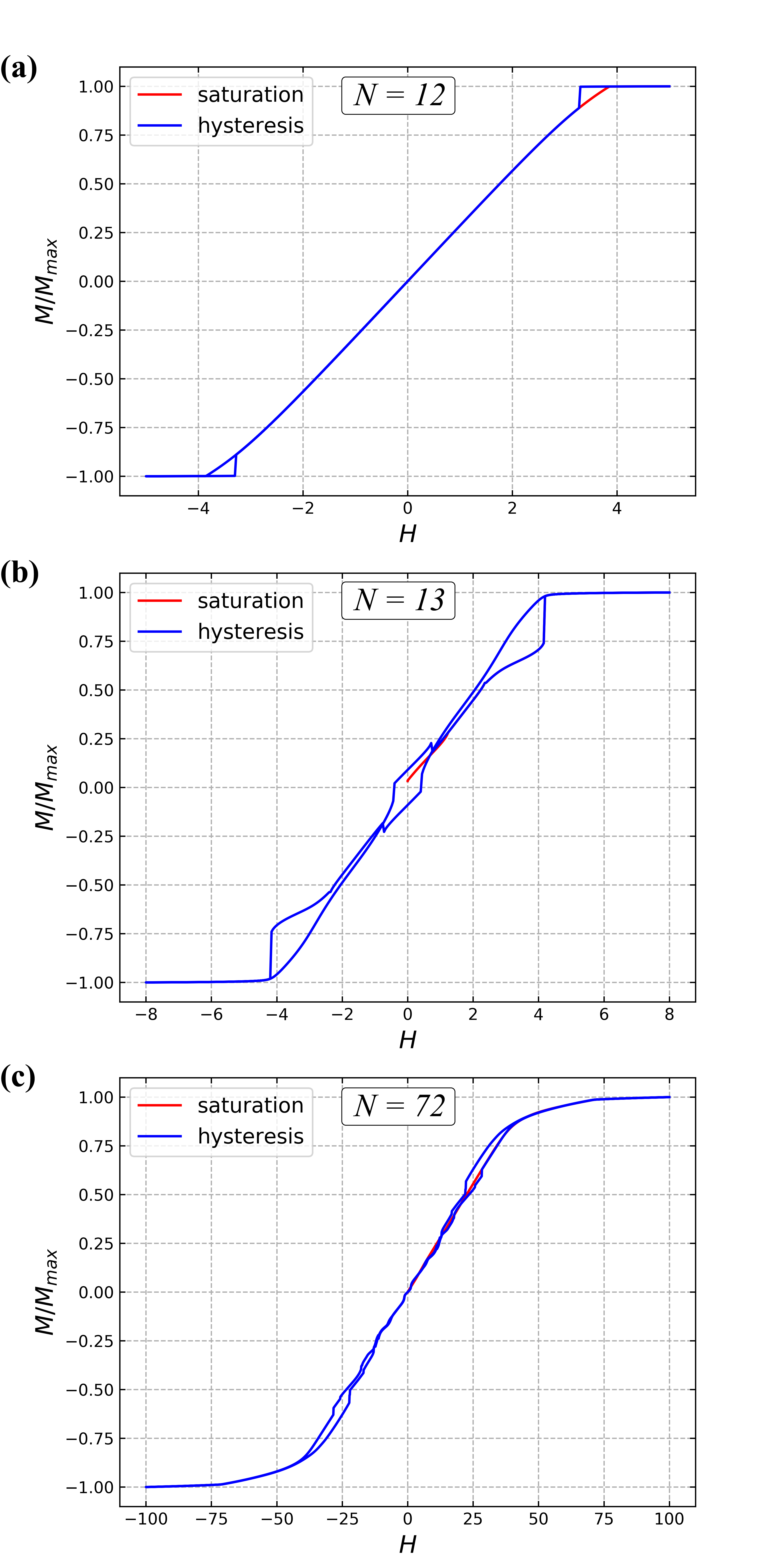} 
	\caption{Hysteresis loops for \textbf{(a)} the ground state for $N=12$, \textbf{(b)} the first excited state for $N=13$ and \textbf{(c)} the fifth excited state for $N=72$. High symmetry states as well as high $N$ configurations show weak hysteresis curves while non-symmetric configurations exhibit greater differences between increasing and decreasing field magnetization curves. After saturation it is possible that the system does not return to the ground state, but a state with nonzero remanent magnetization.}
	\label{fig:hysteresis}
\end{figure}

At the end of this section we look at the magnetization hysteresis loops obtained by decreasing the field amplitude after the system reaches saturation. The results for $N=12$, the first excited state for $N=13$ and fifth excited state for $N=72$ are shown in Fig.~\ref{fig:hysteresis}. The saturated configuration for $N=12$ is metastable which is not the case for $N=13$ and $N=72$ states. We see, however, that saturation curves for both $N=56$ and $N=72$ are different than hysteresis curves at least for small amplitudes of external field which signals that at saturation, the system loses information on the exact initial dipole orientations. In general, hysteresis is observed for all configurations with no symmetry while many symmetric states show no differences between increasing and decreasing field, e.g. ground states for $N=24$ and $N=72$.

\subsection{Freely-rotating sphere}
As discussed, the system's response to external field depends on the direction of this field. We chose the direction of angular momentum as the characteristic direction of order for each configuration, however we saw already for the $N=13$ case that the direction of $\bm{\Gamma}$ also changes during the simulations for configurations with no symmetry. This presents the question of optimal direction of external field at each amplitude -- we seek for the direction that minimizes system energy (Eq.~\ref{hamilton}) at every step. This is equivalent to the case of a freely-rotating sphere which is also more relevant for potential experimental realization. We solve this by also minimizing the energy over the field direction. If we write the interaction Hamiltonian (\ref{hamilton}) in the form $\mathcal{H}_h = -N\bm{H}\cdot\bm{M}$, we see that the optimal external field direction will be parallel to magnetization. With fixed direction of the field, the alignment between field and magnetization is not guaranteed and occurs only at high field amplitudes.

\begin{figure}[!ht] 
	\centering
	\includegraphics[width=\linewidth]{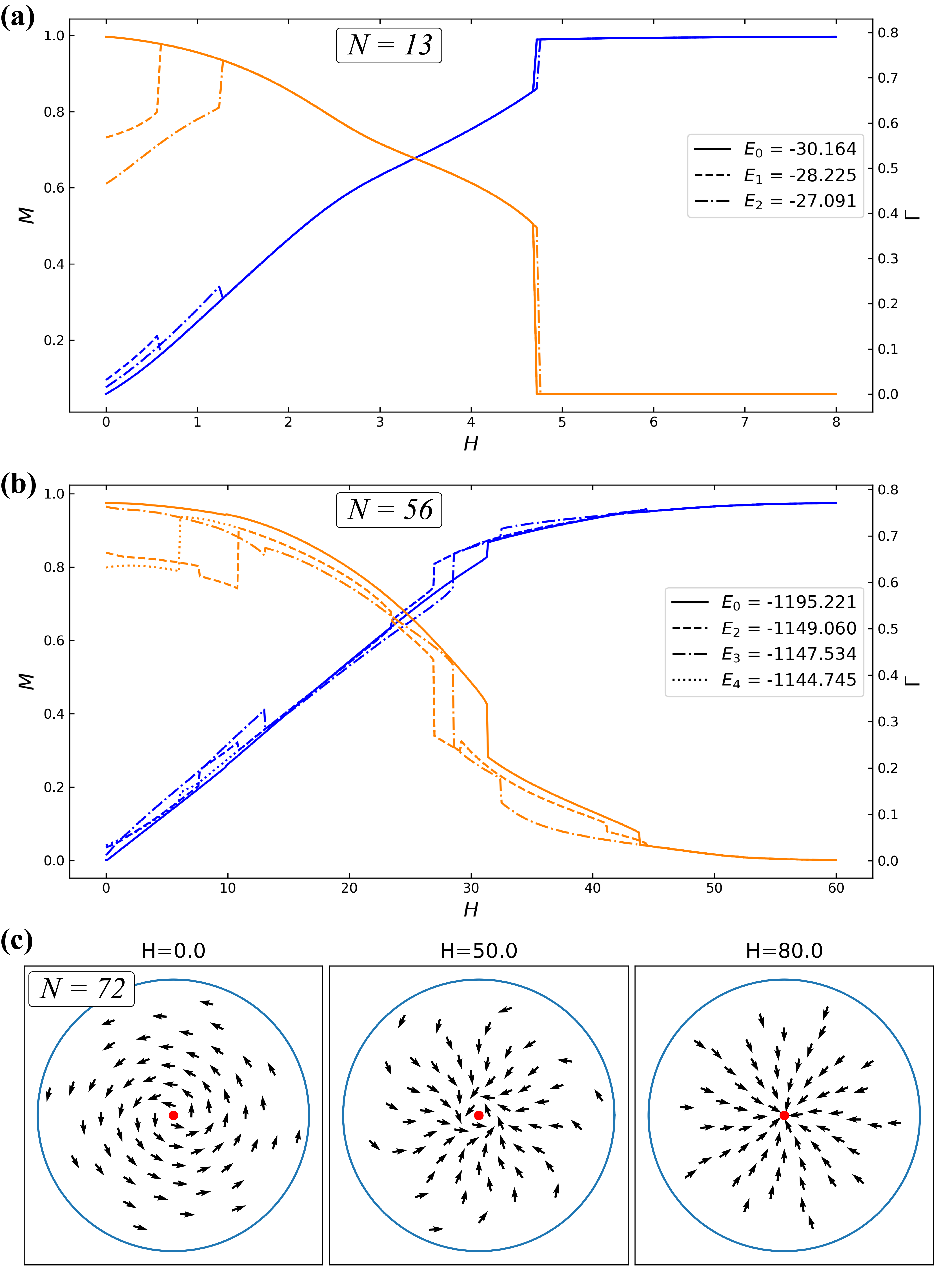}
	\caption{Magnetization and $\bm{\Gamma}$ amplitude of freely rotating sphere in external field for different states of \textbf{(a)} $N=13$ and \textbf{(b)} $N=56$. Compared to the results in Fig.~\ref{fig:zunanje_polje}, the behavior of $N=13$ states is simplified (the number of phase transitions decreases) while the phase transitions for $N=56$ become more pronounced. In both cases the configurations at high field amplitudes are the same for all initial zero field states. \textbf{(c)} Azimuthal plot for the ground state of $N=72$ at different external field amplitudes for the case of the freely-rotating sphere. The plot is centered to the optimal direction of external field (red dot) and we can see that the system symmetry changes from $C_2$ to $C_3$ and ultimately to $C_5$. In the last frame, two dipoles on the symmetry axis are not shown as they are oriented perpendicular to the surface and therefore have no tangent projection.}
	\label{fig:vrteca_krogla}
\end{figure}

Figure~\ref{fig:vrteca_krogla}a shows susceptibilities and angular momentum amplitude for the three states of $N=13$. We notice that the excited states undergo an orientational phase transition to join the configuration with that of the ground state. To see if the states are exactly the same after the junction we could compare the optimal direction of external filed for all configurations and notice that they are the same up to the point group symmetry of positional order. In the saturated configuration, the field is aligned with the symmetry axis of the positional order with dipole configuration also exhibiting $C_{2v}$ symmetry. Different zero field configurations can also stay separated until we reach higher external field amplitudes as shown for the case of $N=56$ in Fig.~\ref{fig:vrteca_krogla}b. Note that the magnetization and angular momentum amplitude curves for both $N=13$ and $N=56$ are different from any of the cases with fixed field direction (Fig.~\ref{fig:vrteca_krogla}b and \ref{fig:vrteca_krogla}d) which means allowing for sphere rotation fundamentally changes dipole states in external field for states with no symmetry. 

For symmetric states where angular momentum is aligned with the rotational symmetry axis one could expect that the external field remains parallel to this axis. This is indeed the case for some states, e.g. $N=12$ where behavior is exactly the same as for the case with fixed field direction, however, optimal field direction can change for some symmetric states which alongside also changes the configuration symmetry. A notable example is the ground state of $N=72$. Figure \ref{fig:vrteca_krogla}c shows azimuthal plot of state configuration at different external field amplitude. The two-fold symmetry of the configuration becomes unstable and the system transitions to a three-fold symmetric state of the first excited state. Finally, at high external field amplitudes, there is another change in optimal field direction where the field aligns with one of the five-fold axes of the positional order. Similar change of symmetry is also observed for the ground state of $N=24$ that transitions from $C_4$ symmetry to $C_3$ at high field amplitudes.

\section{Conclusions}


We explored orientational ordering of point dipoles on a sphere with positional order fixed in the solutions of the Thomson problem by minimizing the system energy. Some parallels to the 2D lattice cases can be drawn, most notably we observe macrovortex ground state that also emerges for the triangular lattice. We expect the macrovortex to be the ground state for other spherical lattices that locally resemble triangular lattice, such as the solutions of the Tammes problem, Thomson problem with generalized power law, and spiral point distributions \cite{Saff1997}.
Configuration symmetries as well as the number of different stable states found, depend strongly on the symmetry of the underlying positional order at each $N$. The symmetry of the dipolar system cannot be higher than the symmetry of the underyling lattice, and in the macrovortex state, can only have a single rotational symmetry axis. We find states belonging to $C_2$, $C_3$ and $C_4$ point groups, however, many macrovortex ground states show no symmetries.
In external field we discover multiple discontinuous orientational phase transitions, especially for less symmetric states. The direction of angular momentum that characterizes ordering of dipoles can also change as we increase external field amplitude. For the case of freely rotating sphere where the external field assumes the direction that minimizes system energy, we find that configurations of different stable states merge with increasing field. In the saturation configuration, the field is aligned with one of the symmetry axes of positional order.

Studying dipolar interactions on a sphere is a step towards understanding and harnessing the role of anisotropic interactions in stability and structure of spherical assemblies. Many biological structures, such as protomers of viral capsids and RNA nanocages, involve electrostatic interactions in addition to chemical bonds and hard core repulsion. These interactions are more complex, and are also often screened by ions in surrounding medium. Generalizations to more complex anisotropic interactions -- screened, quadrupolar or interactions based on empirical models -- are therefore an important open problems for future investigation. The role of thermal fluctuations can further be explored through Monte Carlo simulations. Another potential research direction is the generalization of the problem to allow for movement of dipoles along the surface of the sphere, which corresponds more closely to possible experimental realizations with interacting particles. Based on expected stable structures, predicted from simplified models, bottom-up design of self-assembling nanocontainers can be envisioned, with field-effected orientational transitions and changes in symmetry giving potential for controlled rearrangement or dissolution. 

\begin{acknowledgments}
   We thank Anže Lošdorfer Božič for helpful discussions and suggestions during our research.
  We acknowledge support by Slovenian Research Agency (ARRS) under contracts P1-0099 and J1-9149. The work is associated with COST action CA17139.
\end{acknowledgments}

\bibliographystyle{ieeetr}
\bibliography{bibliography}

\begin{thebibliography}{10}

\bibitem{Thomson}
J.~Thomson, ``On the structure of the atom: an investigation of the stability
  and periods of oscillation of a number of corpuscles arranged at equal
  intervals around the circumference of a circle; with application of the
  results to the theory of atomic structure,'' {\em Lond. Edinb. Dubl. Phil.
  Mag.}, vol.~7, pp.~237--265, Mar. 1904.

\bibitem{Bowick1}
M.~J. Bowick, A.~Cacciuto, D.~R. Nelson, and A.~Travesset, ``Crystalline
  particle packings on a sphere with long-range power-law potentials,'' {\em
  Phys. Rev. B}, vol.~73, p.~024115, Jan. 2006.

\bibitem{Saff1997}
E.~B. Saff and A.~B.~J. Kuilaars, ``Distributing {Many} {Points} on a
  {Sphere},'' {\em The Mathematical Intelligencer}, vol.~19, p.~5, 1997.

\bibitem{Tammes}
P.~Tammes, {\em On the origin of number and arrangement of the places of exit
  on the surface of pollen-grains}.
\newblock PhD thesis, 1930.
\newblock Relation: https://www.rug.nl/ Rights: De Bussy.

\bibitem{Slosar}
A.~Slosar and R.~Podgornik, ``On the connected-charges {Thomson} problem,''
  {\em Europhys. Lett.}, vol.~75, p.~631, 2006.

\bibitem{BowickPRL}
M.~Bowick, A.~Cacciuto, D.~R. Nelson, and A.~Travesset, ``Crystalline {Order}
  on a {Sphere} and the {Generalized} {Thomson} {Problem},'' {\em Phys. Rev.
  Lett.}, vol.~89, p.~237, 2002.

\bibitem{Irvine}
W.~T.~M. Irvine, V.~Vitelli, and P.~M. Chaikin, ``Pleats in crystals on curved
  surfaces,'' {\em Nature}, vol.~468, p.~947, 2010.

\bibitem{Wales}
D.~J. Wales and S.~Ulker, ``Structure and dynamics of spherical crystals
  characterized for the thomson problem,'' {\em Phys. Rev. B}, vol.~74,
  p.~212101, Dec 2006.

\bibitem{Wales2009}
D.~J. Wales, H.~McKay, and E.~L. Altschuler, ``Defect motifs for spherical
  topologies,'' {\em Phys. Rev. B}, vol.~79, p.~1, 2009.

\bibitem{Miller}
W.~L. Miller and A.~Cacciuto, ``Two-dimensional packing of soft particles and
  the soft generalized {Thomson} problem,'' {\em Soft Matter}, vol.~7, p.~7552,
  2011.

\bibitem{Marzec}
C.~Marzec and L.~Day, ``Pattern formation in icosahedral virus capsids: the
  papova viruses and nudaurelia capensis beta virus,'' {\em Biophys. J.},
  vol.~65, pp.~2559--2577, Dec. 1993.

\bibitem{ZandiR_ProcNatlAcadSci101_2004}
R.~Zandi, D.~Reguera, R.~F. Bruinsma, W.~M. Gelbart, and J.~Rudnick, ``From
  {The} {Cover:} {Origin} of icosahedral symmetry in viruses,'' {\em Proc.
  Natl. Acad. Sci.}, vol.~101, p.~15556, 2004.

\bibitem{Kroto}
H.~W. Kroto, J.~R. Heath, S.~C. O'Brien, R.~F. Curl, and R.~E. Smalley, ``C60:
  Buckminsterfullerene,'' {\em Nature}, vol.~318, pp.~162--163, Nov. 1985.

\bibitem{Chevalier}
Y.~Chevalier and M.-A. Bolzinger, ``Emulsions stabilized with solid
  nanoparticles: {Pickering} emulsions,'' {\em Colloid Surface A}, vol.~439,
  p.~23, 2013.

\bibitem{Li2013}
Y.~Li, H.~Miao, H.~Ma, and J.~Z.~Y. Chen, ``Topological defects of tetratic
  liquid-crystal order on a soft spherical surface,'' {\em Soft Matter},
  vol.~9, p.~11461, 2013.

\bibitem{Luca}
J.~D. Luca, S.~B. Rodrigues, and Y.~Levin, ``Electromagnetic instability of the
  {Thomson} problem,'' {\em Europhys. Lett.}, vol.~71, p.~84, 2005.

\bibitem{Sknepnek}
R.~Sknepnek and S.~Henkes, ``Active swarms on a sphere,'' {\em Phys. Rev. E},
  vol.~91, p.~022306, 2015.

\bibitem{Praetorius}
S.~Praetorius, A.~Voigt, R.~Wittkowski, and H.~Löwen, ``Active crystals on a
  sphere,'' {\em Phys. Rev. E}, vol.~97, p.~70, 2018.

\bibitem{Prakash}
S.~Prakash and C.~L. Henley, ``Ordering due to disorder in dipolar magnets on
  two-dimensional lattices,'' {\em Phys. Rev. B}, vol.~42, pp.~6574--6589, Oct.
  1990.

\bibitem{Zimmerman}
G.~O. Zimmerman, A.~K. Ibrahim, and F.~Y. Wu, ``Planar classical dipolar system
  on a honeycomb lattice,'' {\em Phys. Rev. B}, vol.~37, pp.~2059--2065, Feb.
  1988.

\bibitem{Belobrov}
P.~Belobrov, V.~Voevodin, and V.~Ignatchenko, ``Ground state of a dipole in a
  plane rhombic lattice,'' {\em Zh. Eksp. Teor. Fiz.}, vol.~88, pp.~889--893,
  1985.

\bibitem{Bowick2}
M.~J. Bowick and L.~Giomi, ``Two-dimensional matter: order, curvature and
  defects,'' {\em Adv. Phys.}, vol.~58, pp.~449--563, Sept. 2009.

\end{thebibliography}

\end{document}